\documentclass[aps,pra,twocolumn,superscriptaddress,showpacs,preprintnumbers]{revtex4}
\usepackage{makeidx}
\usepackage{amsmath}
\usepackage{graphicx}
\usepackage{dcolumn}
\usepackage{bm}
\usepackage{amssymb}
\usepackage{color}
\usepackage{subfig}
\input{tcilatex}
\begin{document}

\title{Influence of the surface band structure on electron emission spectra
from metal surfaces}
\author{C. D. Archubi }
\affiliation{Instituto de Astronom\'{\i}a y F\'{\i}sica del Espacio (CONICET-UBA),
casilla de correo 67, sucursal 28, C1428EGA, Buenos Aires, Argentina}
\author{M. N. Faraggi }
\affiliation{Instituto de Astronom\'{\i}a y F\'{\i}sica del Espacio (CONICET-UBA),
casilla de correo 67, sucursal 28, C1428EGA, Buenos Aires, Argentina}
\author{V. M. Silkin}
\affiliation{Donostia International Physics Center (DIPC), 20018 San Sebasti\'{a}n, Spain}
\affiliation{Depto. de F\'{\i}sica de Materiales, Facultad de Ciencias Qu\'{\i}micas,
Universidad del Pa\'{\i}s Vasco, Apdo. 1072, 20080 San Sebasti\'an, Spain}
\affiliation{IKERBASQUE, Basque Foundation for Science, 48011 Bilbao, Spain}
\author{M. S. Gravielle}
\affiliation{Instituto de Astronom\'{\i}a y F\'{\i}sica del Espacio (CONICET-UBA),
casilla de correo 67, sucursal 28, C1428EGA, Buenos Aires, Argentina}
\date{\today }

\begin{abstract}
Electron distributions produced by grazing impact of fast protons on
Mg(0001), Cu(111), Ag(111) and Au(111) surfaces are investigated, focusing
on the effects of the electronic band structure. The process is described
within the Band-Structure-Based approximation, which is a perturbative
method that includes an accurate representation of the electron-surface
interaction, incorporating information of the electronic band structure of
the solid. For all the studied surfaces, the presence of partially occupied
surface electronic states produces noticeable structures in double
differential - energy- and angle- resolved - electron emission probabilities
from the valence band. These structures remain visible in electron emission
spectra after adding contributions coming from core electrons, which might
make it possible their experimental detection.
\end{abstract}

\maketitle

\section{Introduction}

When a charged particle moves parallel to a metal surface different
processes may take place \cite{araussr97,wipr02} like energy loss of the
ion, charge transfer between particle and metal, inner-shell and
valence-band electronic excitations and secondary electron emission. In
particular, the interaction of the projectile with valence electrons of the
solid produces a rearrangement of the charges inside the material, a dynamic
process that gives rise to one-particle electronic transitions
(electron-hole excitations) and collective excitation (plasmon) processes.
For decades these processes were studied using simple representations of the
unperturbed electronic states of the surface, while the emphasis was set on
the description of the response of the many-electron system to the external
perturbation. However, recent experimental and theoretical works showed that
the band structure of different metal surfaces plays an important role in
projectile induced processes \cite{hewiprl00,hewifd00,bomepra03,Silkin2003,
Silkin2005, Silkin2007, Silkin2008, Vattuone, Diaconescu, Park,
Jahn,govaprb11,siruss07,pirinnl12,sipinimb11}. In surfaces like Cu(111),
Ag(111) and Au(111) it was found that the energy lost by charged particles
moving at long distances from the surface is dominated by electron-hole
excitations involving partially occupied surface states \cite{Silkin2005}.
Also in the case of Mg(0001) the finite width of the surface plasmon
modifies the behavior of the energy loss at long distances from the surface,
being relevant for the study of the interaction mechanisms between charged
particles and the internal walls of microcapilars \cite{Silkin2008, Tokesi}.

In the above mentioned articles, the influence of the surface band structure
on energy loss processes was investigated making use of the dielectric
formalism. This formalism involves a quantum calculation of the surface
response function but describes the stopping of the incident particle by
means of classical electromagnetism laws. Here we are interested in studying
individual electronic transitions induced by the projectile. In particular,
the work focuses on angle- and energy- resolved electron emission
distributions produced by grazing impact of swift protons on metal surfaces.
Such spectra are expected to provide detailed information on the electronic
characteristics of the target surface.

To describe the electron emission process from the valence band we employ
the Band-Structure-Based (BSB) approximation \cite{Marisa,Archubi}, which is
derived within the framework of the binary collisional formalism by
including a precise representation of the surface interaction. For every
individual electronic excitation, the BSB transition matrix is evaluated
making use of the electronic states corresponding to the model potential of
Ref. \cite{ChulkovSilkin}, which incorporates effects of the band structure
of the metal. This potential has been successfully used in several areas
\cite{Silkin2003, Silkin2005, Silkin2007, Silkin2008, Marisa,
Faraggi2005,ChulkovSilkin,Archubi, CarlosRios}, reproducing properly the
projected energy gap and the energies of the surface and first image states.
Within the BSB method, the dynamic response of the medium to the incident
charge is obtained from the unperturbed electronic wave functions by using
the linear response theory.

In a previous work \cite{Archubi} the BSB \ approximation was applied to
study electron emission induced by grazing incidence of fast protons on a
Be(0001) surface. It was found that the distribution of ejected electrons
presents prominent signs of the surface band structure, with pronounced
shoulders due to the contribution of partially occupied surface electronic
states (SESs). In this paper the research is extended to consider electron
emission from Cu(111), Ag(111), Au(111) and Mg(0001) surfaces, for which it
is foreseeable that band structure effects leave footprints on the electron
emission spectra like the ones observed in stopping processes. Contributions
from the inner shells of surface atoms, calculated with the
Continnum-distorted-wave-Eikonal-initial-state (CDW-EIS) approximation, are
also included in the spectra in order to determine the energy and angular
range where band structure effects might be experimentally detected. In
addition, with the aim of investigating the dependence on the incidence
conditions, the influence of the projectile trajectory is analyzed.

The paper is organized as follows. In Sec. II we summarize the theoretical
model, results are presented and discussed in Sec. III, and Sec. IV contains
our conclusions. Atomic units are used unless otherwise stated.

\section{Theoretical method}

We consider a projectile $P$ that impinges grazingly on a metal surface,
inducing the one-electron transition $i\rightarrow f$, where the state $i$\
belongs to the valence band while the state $f$ \ lays in the continuum.
Within the binary collisional formalism, the corresponding transition
probability per unit path reads \cite{Gravielle}:

\begin{equation}
P_{if}\left( Z\right) =\frac{2\pi }{v_{s}}\delta \left( \Delta \right)
\left\vert T_{if}\right\vert ^{2},  \label{Pif}
\end{equation}%
where $Z$ is the projectile distance to the surface, $v_{s}$ is the
component of the projectile velocity parallel to the surface plane, and the
Dirac delta function $\delta \left( \Delta \right) \ $expresses the energy
conservation, with
\begin{equation}
\Delta =\vec{v}_{s}.\left( \vec{k}_{fs}-\vec{k}_{is}\right) -\left(
E_{f}-E_{i}\right) ,  \label{conservation}
\end{equation}%
$\vec{k}_{is}$ ($\vec{k}_{fs}$) the initial (final) electron momentum
parallel to the surface, and $E_{i}$ ($E_{f}$) the initial (final) electron
energy. In Eq. (\ref{Pif} ) $T_{if}$ represents the T-matrix element, which
is evaluated within a first-order perturbation theory as:

\begin{equation}
T_{if}=\left\langle \Phi _{f}\left\vert V\right\vert \Phi _{i}\right\rangle ,
\label{T}
\end{equation}%
where $\Phi _{i}(\Phi _{f})\ $is the initial (final) unperturbed electronic
state, evaluated with the BSB model, and $V$ denotes the perturbative
potential produced by the external charge.

By assuming translational invariance in the plane parallel to the surface,
the BSB unperturbed states, \ $\Phi _{i}=\Phi _{\vec{k}_{is},n_{i}}$\ \ and $%
\Phi _{f}=\Phi _{\vec{k}_{fs},n_{f}}$, are expressed as:

\begin{equation}
\Phi _{\vec{k}_{s},n}\left( \vec{r}\right) =\frac{1}{2\pi }\exp \left( i\vec{%
k}_{s}.\vec{r}_{s}\right) \phi _{n}\left( z\right) ,
\label{unperturbed states}
\end{equation}%
where $\vec{r}=$ ($\vec{r}_{s},z)$ is the position vector of the active
electron, with $\vec{r}_{s}$ and $z$ being the components of $\vec{r}$ \
parallel and perpendicular, respectively, to the surface plane. The function
$\phi _{n}(z)$ represents the eigenfunction of the one-dimensional Schr\"{o}%
dinger equation associated with the surface potential of Ref. \cite%
{ChulkovSilkin} with eigenenergy $\varepsilon _{n}.$ By using slab geometry,
the eigenfunctions $\phi _{n}(z)$ can be classified as symmetric (S) or
antisymmetric (A) according to the reflection symmetry properties with
respect to the center of the slab. They read:

\begin{equation}
\phi _{n}^{(S)}\left( z\right) =\frac{1}{\sqrt{L}}c_{n}^{(S)}(0)+\frac{1}{%
\sqrt{L}}\overset{N}{\underset{l=1}{\sum }}c_{n}^{(S)}\left( l\right) \cos
\left( \frac{2\pi l}{L}\widetilde{z}\right) ,  \label{Slab2}
\end{equation}%
\begin{equation}
\phi _{n}^{(A)}\left( z\right) =\frac{1}{\sqrt{L}}\overset{N}{\underset{l=1}{%
\sum }}c_{n}^{(A)}\left( l\right) \sin \left( \frac{2\pi l}{L}\widetilde{z}%
\right) ,  \label{Slab}
\end{equation}
where $L$ is the length of the unit cell, $2N+1$ is the number of basis
functions, and the coefficients $c_{n}^{(j)}\left( l\right) $, $j=S,A$, are
obtained numerically \cite{Silkin2005}. The coordinate $\widetilde{z}%
=z+d_{s}\ $represents the normal distance measured with respect to the
center of the slab, which is placed at a distance $d_{s}$ from the surface
plane.

In Eq. (\ref{T}) the potential \ $V$ is expressed as $V=V_{Pe}+V_{ind}$,
where $V_{Pe}=-Z_{P}/r_{P}$ denotes the Coulomb projectile potential, with $%
Z_{P}$ the projectile charge and $r_{P}$ the electron-projectile distance,
and $V_{ind}$\ represents the surface potential induced by the incident ion
moving at a distance $Z$ from the surface plane.\ The potential $V_{ind}$ \
is obtained from the two-dimensional Fourier transform of the
density-density response function, which is calculated within the linear
response theory by employing the BSB unperturbed electronic states given by
Eq. (\ref{unperturbed states}) \cite{egprl83,Silkin2005}.

The differential electron transition probability from the valence band$\ $to
a given final state $f\ $\ with momentum $\vec{k}_{f}$,$\ dP_{vb}/d\vec{k}%
_{f}$, is derived by integrating Eq. (\ref{Pif}) along the classical
projectile path, after adding the contributions coming from the different
initial states. That is,

\begin{equation}
\frac{dP_{vb}}{d\vec{k}_{f}}=\underset{-\infty }{\overset{+\infty }{\int }}dX%
\left[ \sum\limits_{i}\rho _{e}\ \Theta \left( -E_{W}-E_{i}\right)
P_{if}\left( Z\left( X\right) \right) \right] ,  \label{dPdk}
\end{equation}%
where $Z\left( X\right) $ is the projectile trajectory, with \ $X$ the
coordinate along the incidence direction, $\rho _{e}=2$ takes into account
the spin states, \ and the unitary Heaviside function $\Theta $ restricts
the initial states to those contained inside the Fermi sphere, with $E_{W}$
the work function. Notice that within the BSB model the final electronic
states $\Phi _{f}$ present a well defined momentum only in the direction
parallel to the surface plane. Then, in order to determine $\vec{k}_{f}=(%
\vec{k}_{fs},k_{fz})$ it is necessary to define an \textit{effective}
electron momentum perpendicular to the surface as $k_{fz}=\sqrt{2\varepsilon
_{n_{f}}}$, where $\varepsilon _{n_{f}}$ is the eigenenergy associated with
the final one-dimensional wave function $\phi _{n_{f}}\left( z\right) $.
More details of the BSB approximation can be found in Ref. \cite{Marisa}.

\section{Results}

We apply the BSB approximation to investigate electron emission spectra
produced by grazing scattering of protons from different metal surfaces -
Mg(0001), Cu(111), Ag(111) and Au(111) - considering incidence velocities in
the high energy range, i.e. ranging from 1.5 a.u. to 3.5 a.u. At this impact
energies electron capture processes are negligible and protons can be
treated as bare ions along the whole path \cite{GravielleMiraglia1994}. To
evaluate the classical projectile trajectory we employed the
Ziegler-Biersack-Littmark potential \cite{ZBL} plus the BSB induced
potential, which was derived in a consistent way by using the linear
response theory \cite{Silkin2005}.

The differential probability of electron emission from the valence band, $%
dP_{vb}/d\vec{k}_{f}$, was obtained from Eq. (\ref{dPdk}) by interpolating
the $P_{if}$ function, given by Eq. (\ref{Pif}), from data corresponding to $%
24$ different $Z$ distances. The $T_{if}$ elements were calculated by using
the BSB wave functions of Eq. (\ref{unperturbed states}), where the
one-dimensional wave functions $\phi _{n}$ were derived by following the
same procedure as in Refs. \cite{Marisa, Faraggi2005, Archubi}. A slab
formed by $40$ atomic layers was used for the different targets, while the
number of layers associated with the vacuum was chosen as $20$ for Cu, Ag
and Au, and $10$ for the case of Mg. The length of the unit cell ($L$)
considered for every material varies between $250$ $\mathrm{{\mathring{A}}}$
and $300$ $\mathrm{{\mathring{A}}}$ and the sums in Eqs. (\ref{Slab2}) and (%
\ref{Slab}) run up to $N=150$. \ The Fermi energy level was determined from
the work function values reported in Ref. \cite{ChulkovSilkin}. Notice that
all the studied surfaces present SESs that are partially occupied (see Fig.
1), with energies varying from $\varepsilon _{SES}=-4.62$ eV for silver to $%
\varepsilon _{SES}=-6.02$ eV for gold, the energy values being measured with
respect to the vacuum level.

Within the BSB model two\ $\phi _{n_{f}}\left( z\right) $ functions - the
symmetric one and the antisymmetric one - are associated with the same
energy $\varepsilon _{n_{f}}$ in the thick slab limit. This representation
does not allow to distinguish electrons emitted inside the solid from those
ejected towards the vacuum semi-space. Then, as a first estimate we
considered that ionized electrons emitted to the vacuum region represent
approximately a $50\%$ of the total ionized electrons from the valence band
\cite{Marisa, Faraggi2005}.

With the aim of presenting an overall scenario of the influence of the
electronic band structure for the different materials, in Fig. 2 we show $%
dP_{vb}/d\vec{k}_{f}$, as a function of the electron energy, for protons
impinging on \ Mg, Cu, Ag and Au surfaces with the incidence velocity $v=2\;
$a.u. and the glancing angle $\alpha =0.1^{\circ }$. In the figure, results
for the ejection angle $\theta _{e}=30^{\circ }$, measured with respect to
the surface in the scattering plane, are displayed by using the same scale
for all the cases. These spectra show the typical double-peaked structure
associated with soft and binary single-particle collisions, respectively,
with conduction electrons \cite{Gravielle00}. But in addition to such
structures, the BSB curves exhibit a noticeable superimposed bulge in the
high electron energy region, which disappears completely when partially
occupied SESs are not included in the calculations. The shape and size of
this elevation depends on the material, looking like a large shoulder for
Mg, Cu and Au, while for Ag the structure resembles a small protuberance of
the electron distribution. We found that the contribution coming from SESs,
also shown in the figure, is responsible for the superimposed structure of
the electron spectrum. This is due to the fact that SESs present highly
peaked electron densities near the surface, as shown in Fig. 1, which favors
the electron emission when the projectile moves far from the surface plane.
As a result, the SES contribution is more relevant when, in the selvage
region, differences between the electronic density associated with the SES
and those corresponding to other occupied electronic states are larger,\ as
it happens for the copper, silver and gold surfaces, for which the
differences rise up to a factor larger than $70$. On the other hand, when
the electron-surface interaction is represented by a finite step potential (%
\textit{jellium} model) \cite{Gravielle}, without taking into account the
electronic band structure, the electron emission distributions display a
smooth behavior as a function of the electron energy, also observed for
Be(0001) surfaces \cite{Archubi}.

In order to analyze the angular dependence of the SES contribution,
differential probabilities for electron emission from the valence band of
Cu(111) are plotted in Fig. 3, as a function of the electron energy, for
emission angles $\theta _{e}$ ranging from $20^{\circ }$ to $70^{\circ }$.\
For the smaller $\theta _{e}$ values the SES bulge is placed around the
energy of the binary maximum and when the ejection angle increases, the SES
structure moves gradually to the low energy region. Simultaneously, an
additional SES peak arises at low energies, which ends\ up joined to the SES
shoulder for $\theta _{e}\succeq 60^{\circ }$. This behavior is ruled by the
energy conservation imposed by the delta function of Eq. (\ref{conservation}%
), which determines the maximum ($k_{SES}^{(+)}$ ) and minimum\ ($%
k_{SES}^{(-)}$) final momenta reached by transitions from occupied SESs.
These momenta verify%
\begin{equation}
k_{SES}^{(\pm )}=v_{s}\cos \theta _{e}+\sqrt{R_{\pm }^{2}-v_{s}^{2}\sin
^{2}\theta _{e}},  \label{k}
\end{equation}%
where $R_{\pm }^{2}=$ $(k_{SES}\pm v_{s})^{2}+2\varepsilon _{SES}$, with $%
k_{SES}=$ $\sqrt{-2(E_{W}+\varepsilon _{SES})}$. The right hand side of Eq (%
\ref{k}) decreases when $\theta _{e}$ increases and, consequently, the
maximum and minimum energies associated with SES emission shift to lower
values.

Even though for all the considered surfaces, SES structures are clearly
visible in electron emission spectra from the valence band, in experimental
electron distributions \cite{Grizzi} there is another source of ejected
electrons - the inner shells of surface atoms \ - which might hide SES
effects. To evaluate the inner-shell emission yield we employ a
semiclassical formalism \cite{Gravielle00} that describes the multiple
collisions of the incident ion with the surface atoms as single encounters
with outermost atoms along the projectile path. In the model the core
emission probability per unit path is expressed in terms of atomic
probabilities, which are evaluated within the CDW-EIS approximation. The
CDW-EIS approach is a distorted wave method that accounts for the proper
asymptotic conditions \cite{Crothers}, including the distortion produced by
the projectile in both the initial and final states. In the calculation of
the atomic probability we have taken into consideration the full dependency
of the CDW-EIS transition amplitude on the impact parameter, that is, not
only on the modulus of the impact parameter but also on its direction.

Total emission probabilities obtained as the sum of valence and core
contributions are plotted in Fig. 4, together with the partial valence and
inner-shell distributions, for electron emission from a Cu(111) surface with
different ejection angles. The core emission probability from the copper
surface was evaluated by including the 3d- level only, since contributions
coming from deeper shells are expected to be negligible. To represent the 3d
initial state of Cu we used the Hartree-Fock wave function corresponding to
the Cu$^{+}$ ion \cite{Clementi-Roetti}, considering that the outermost
electron was assigned to the valence band. The final continuum state,
associated with the electron ejected from the 3d- level, was described as a
Coulomb wave function with an effective charge satisfying the initial
binding energy. The figure shows that core electrons give rise to a
probability that decreases evenly when the electron energy increases. It
represents the main contribution to the electron emission spectra in almost
the whole energy and angular range. However, signatures of the SES emission
are still present in total electron distributions. At intermediate ejection
angles, the SES structure of the valence-band distribution is also visible,
albeit weakened, in the total emission spectrum, producing an increase of
total probability around the SES position that varies between $20\%$ and $%
35\%$ approximately. In turn, for $\theta _{e}\succeq 60^{\circ }$ SES
effects are reflected as a change in the slope of the electron energy
distribution. Notice that under the condition of grazing incidence,
transport effects are expected to play a minor role \cite{Kimura98}, at
least for the electron ejection angles considered here, and consequently,
present theoretical spectra might be directly compared with experimental
data.

When the atomic number of surface atoms increases, the inner-shell
contribution to the electron emission process augments, as it happens for Ag
and Au surfaces, displayed in Figs. 5 and 6, respectively. For silver,
inner-shell emission corresponding to the 4d- level was calculated by using
the Hartree-Fock wave function of Ag$^{+}$ \cite{Clementi-Roetti}, while for
gold the core contribution from the 5d- level was evaluated by employing the
relativistic wave function of Ref. \cite{Montanari}.\ In both cases, at the
smaller ejection angles, core emission represents the dominant mechanism
that partially conceals surface band structure effects in electron emission
spectra. For Ag(111), a small SES structure is perceivable in total electron
distributions for ejection angles lower than $70^{\circ }$, while for larger
angles the SES contribution produces a change of the slope at low electron
energies, also observed for Cu surfaces. But for Au(111), despite the
remarkable contribution of the electron emission from partially occupied
SESs, at intermediate angles its effects are almost completely covered by
the inner-shell contribution, being only noticeable for $\theta _{e}\succeq
70^{\circ }$.

On the contrary, for Mg(0001) valence-band electrons provide the main
contribution to the electron emission process, as shown in Fig. 7. In this
case, inner-shell emission from the L- shell of Mg cores \cite{aclaracion}
is more than one order of magnitude lower than the valence-band
contribution, except for the lower $\theta _{e}$ values, precisely in the
energy region where electron emission from the valence band is forbidden as
a result of energy conservation (Eq. (\ref{conservation})). Then, although
SES effects for Mg(0001) are weaker than for the previous surfaces, they are
appreciable in total electron distributions for a wide range of ejection
angles.

Finally, we address the study of the influence of the incidence conditions
on \ SES effects, taking as prototype the Cu(111) surface. In Fig. 8.a we
plot $dP_{vb}/d\vec{k}_{f}$ for protons impinging grazingly on Cu(111) with
velocities ranging from $v=1.5\;$a.u. to $v=3.5\;$a.u. The ejection angle
was chosen as $\theta _{e}=30^{\circ }$. \ We found that the SES shoulder
shifts to higher electron energies as the projectile velocity increases, in
accord with Eq. (\ref{k}), but its relative contribution changes moderately
as the velocity varies. However, when the incidence angle $\alpha $ augments
keeping the velocity as a constant, the SES structure becomes smaller,
producing only a smooth shoulder in the electron emission probability for $%
\alpha =0.75^{\circ }$, as shown in Fig. 8.b. This behavior is due to the
fact that large incidence angles allow protons to reach distances closer to
the surface, inducing a strong electron emission from different occupied
electronic states. But when projectiles move far away from the surface
plane, as it happens for the lower $\alpha $ values, only SES electrons are
strongly affected by the external perturbation, giving rise to a remarkable
SES contribution. Similar behavior was also observed for the other surfaces.
Then, for the studied surfaces SESs might be experimentally probed by proton
impact with glancing angles.

\section{Conclusion}

Electron emission spectra \ produced by grazing incidence of protons\ on
Mg(0001), Ag(111), Cu(111), and Au(111) surfaces have been studied,
including valence-band and inner-shell contributions. For all the considered
surfaces, BSB differential emission probabilities from the valence band
display noticeable structures due to the presence of partially occupied
SESs. Such structures are related to the high localization \ of the
electronic density of the SES around the selvage region, which promotes the
electron emission process for projectiles moving outside the solid.
Consequently, SES structures are more pronounced \ for glancing incidence
angles, moving to higher energies as the incidence velocity increases.

We found that SES structures are clearly visible in total emission spectra
for Mg, Cu and Ag surfaces. But \ for Au, band structure effects become
softened and even completely covered by the inner-shell emission, being only
detectable at large ejection angles. We hope the present work will prompt
experimental research on the subject.

\acknowledgments This work was partially supported by the Universidad de
Buenos Aires, the Agencia Nacional de Promoci\'{o}n Cient\'{\i}fica y Tecnol%
\'{o}gica of Argentina, the Consejo Nacional de Investigaciones Cient\'{\i}%
ficas y T\'{e}cnicas (CONICET)

\begin{figure}[h!]
\begin{center}
\includegraphics*[width=9cm]{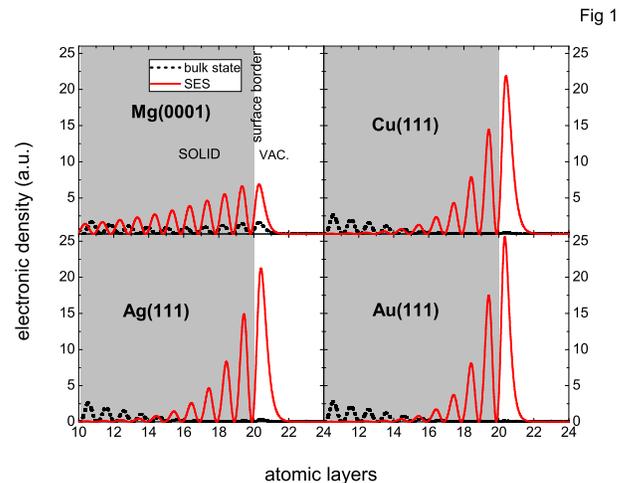}
\end{center}
\caption{(Color online) Comparison between the electronic density of
the SES and the one corresponding to a bulk-like state with a close
energy value for a) Mg(0001), b) Cu(111), c) Ag(111), and d)
Au(111).} \label{fig1}
\end{figure}

\begin{figure}[h!]
\begin{center}
\includegraphics*[width=9cm]{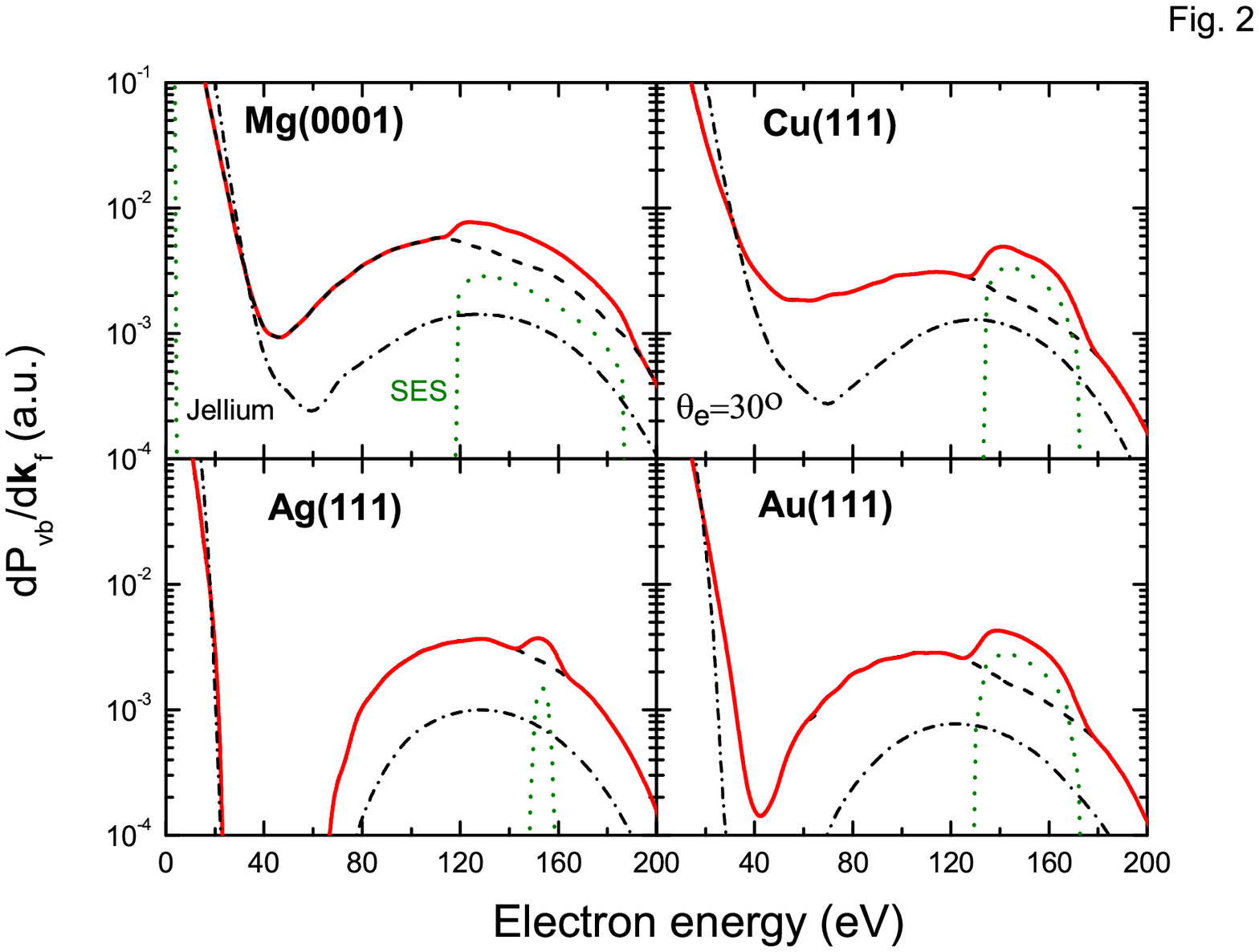}
\end{center}
\caption{(Color online) Differential probability of electron
emission from the valence band, as a function of the electron
energy, for 100 keV protons impinging with the glancing angle
$\protect\alpha =0.1^{o}$ on a) Mg(0001),
b) Cu(111), c) Ag(111), and d) Au(111). The electron ejection angle is: $%
\protect\theta _{e}=30^{o}$, measured with respect to the surface in the
scattering plane. Solid red line, BSB results including SES contribution;
dashed black line, BSB results without the SES contribution; dotted green
line, SES contribution; dash-dotted black line, results obtained within the
jellium model \protect\cite{Gravielle}.}
\label{fig2}
\end{figure}

\begin{figure}[h!]
\begin{center}
\includegraphics*[width=9cm]{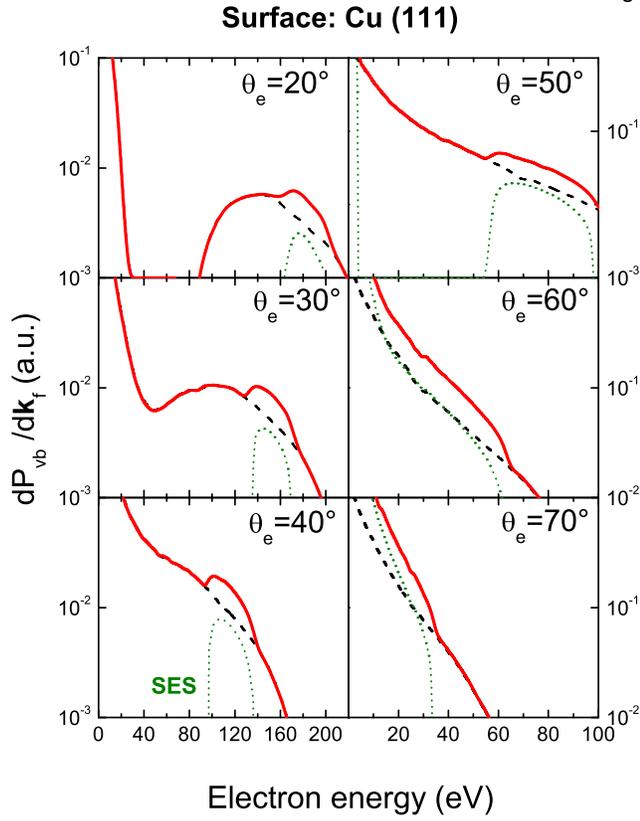}
\end{center}
\caption{(Color online) Influence of the SES in differential
electron emission probabilities for 100 keV protons impinging on a
Cu(111) surface with an incidence angle $\protect\alpha =0.5^{o}$.
The electron ejection
angles are $\protect\theta _{e}=20^{o}$, $30^{o}$, $40^{o}$, $50^{o}$, $%
60^{o}$ and $70^{o}$, respectively, all of them measured with respect to the
surface in the scattering plane. Solid red line, BSB electron emission
spectrum; dashed black line, BSB electron emission spectrum without
including the SES contribution; dotted green line, SES contribution.}
\label{fig3}
\end{figure}

\begin{figure}[h!]
\begin{center}
\includegraphics*[width=9cm]{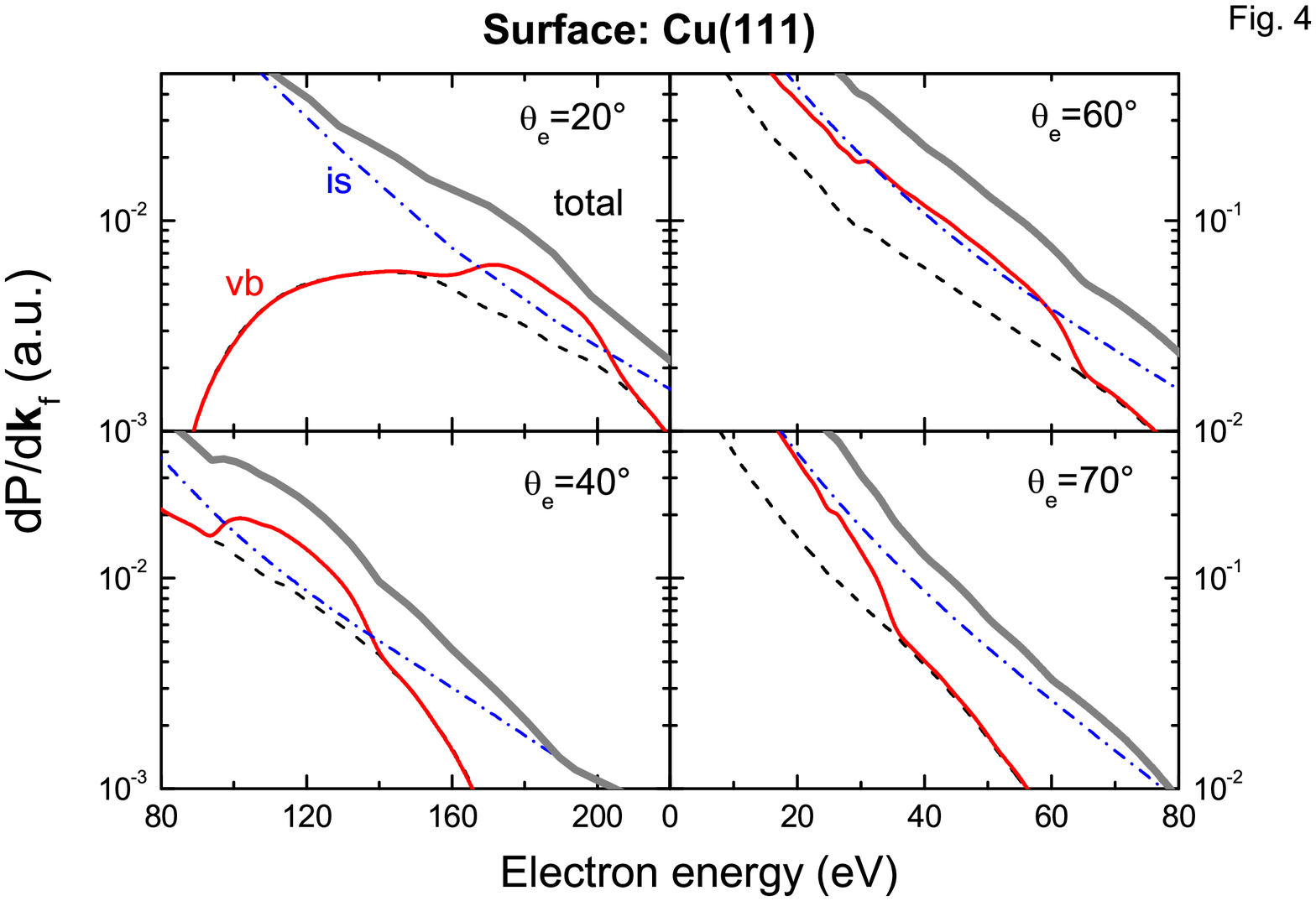}
\end{center}
\caption{(Color online) Similar to Fig. 3 for total emission
probability, including valence-band and inner-shell contributions.
The ejection angles are $\protect\theta _{e}=20^{o}$, $40^{o}$,
$60^{o}$, and $70^{o}$, respectively. Solid thick gray line, total
emission probability obtained by adding valence band and core
contributions, as explained in the text; solid thin red line, BSB
valence emission probability; dashed black line, BSB probability
without including the SES contribution; dot-dashed blue line,
inner-shell emission probability. }
\label{fig4}
\end{figure}

\begin{figure}[h!]
\begin{center}
\includegraphics*[width=9cm]{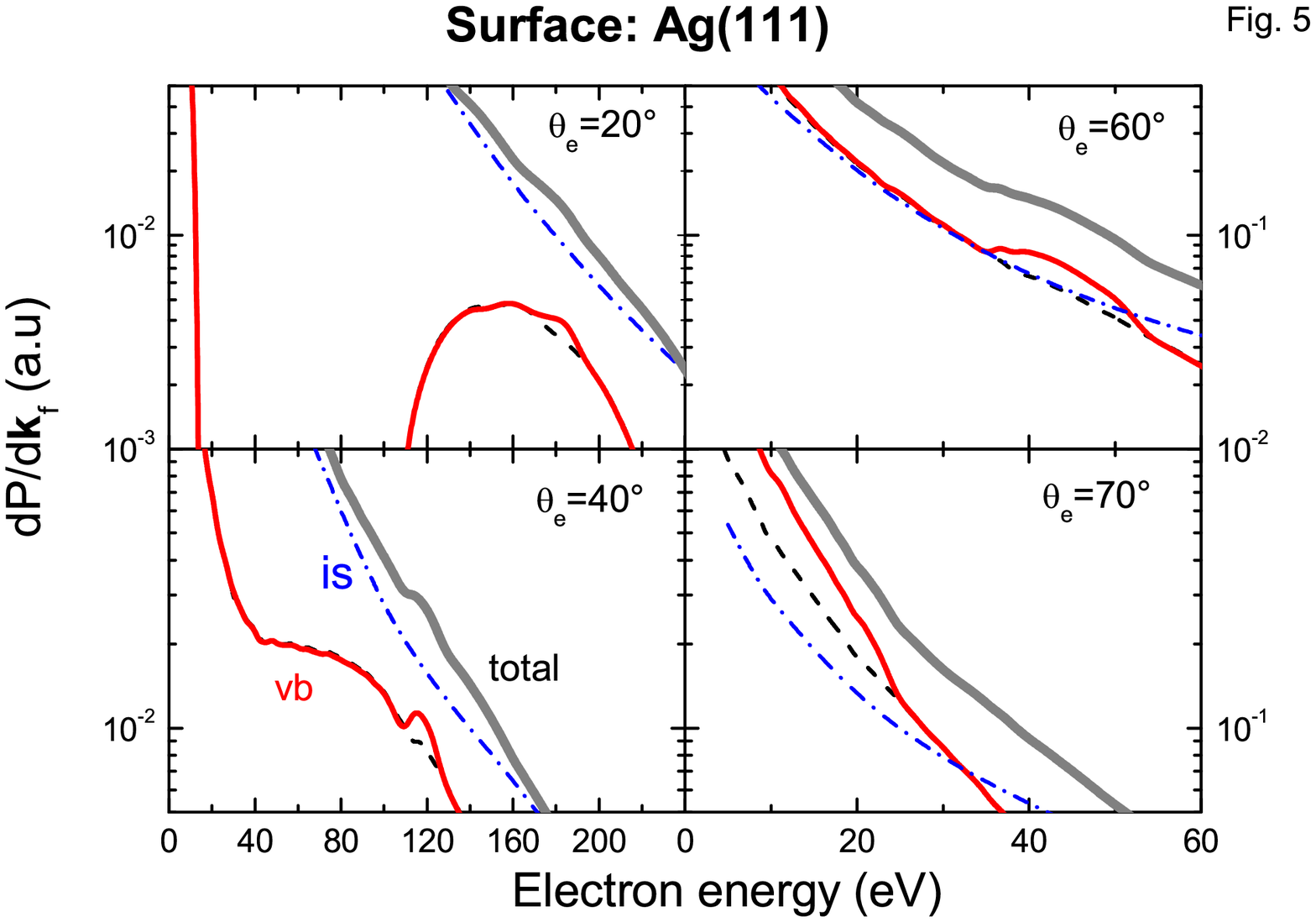}
\end{center}
\caption{(Color online) Similar to Fig 4 for a Ag(111) surface. }
\label{fig5}
\end{figure}


\begin{figure}[h!]
\begin{center}
\includegraphics*[width=9cm]{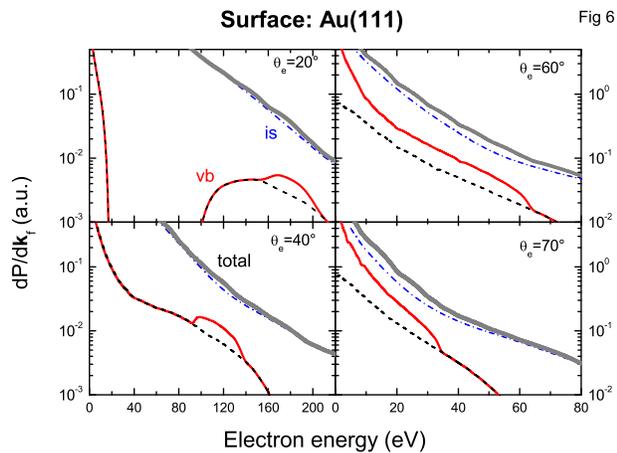}
\end{center}
\caption{(Color online) Similar to Fig 4 for a Au(111) surface.}
\label{fig6}
\end{figure}


\begin{figure}[h!]
\begin{center}
\includegraphics*[width=9cm]{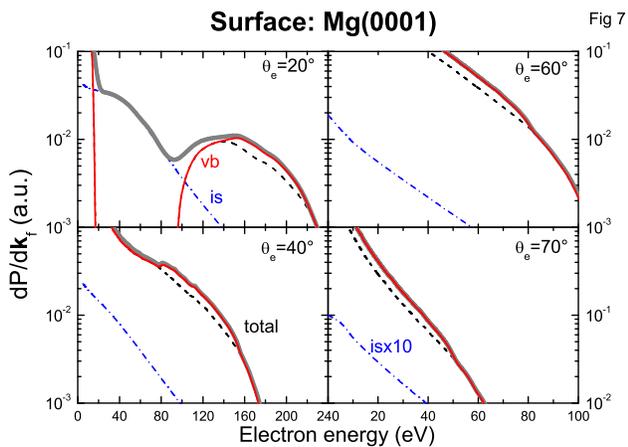}
\end{center}
\caption{(Color online) Similar to Fig 4 for a Mg(0001) surface.}
\label{fig7}
\end{figure}


\begin{figure}[h!]
\begin{center}
\includegraphics*[width=9cm]{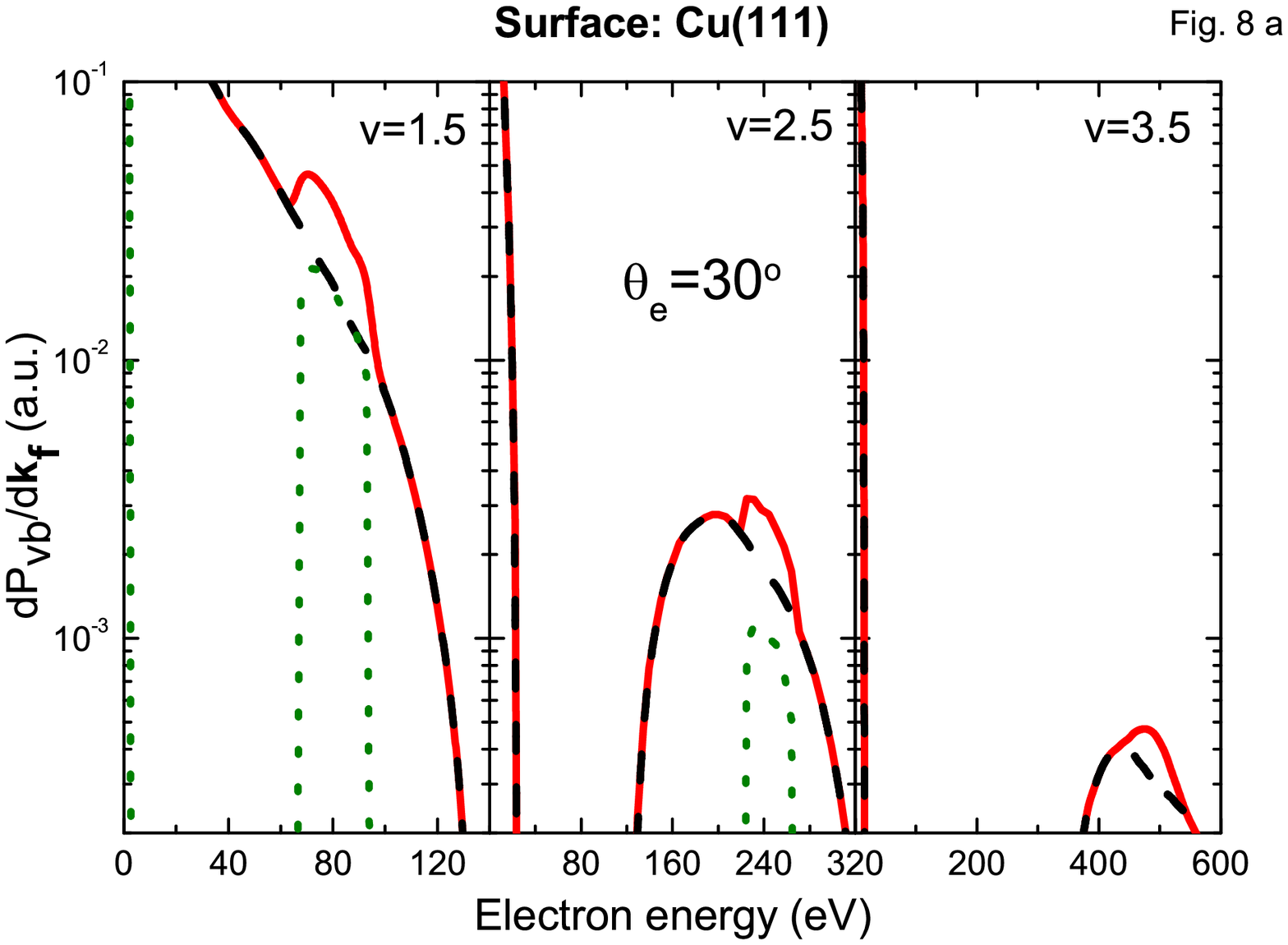}
\includegraphics*[width=9cm]{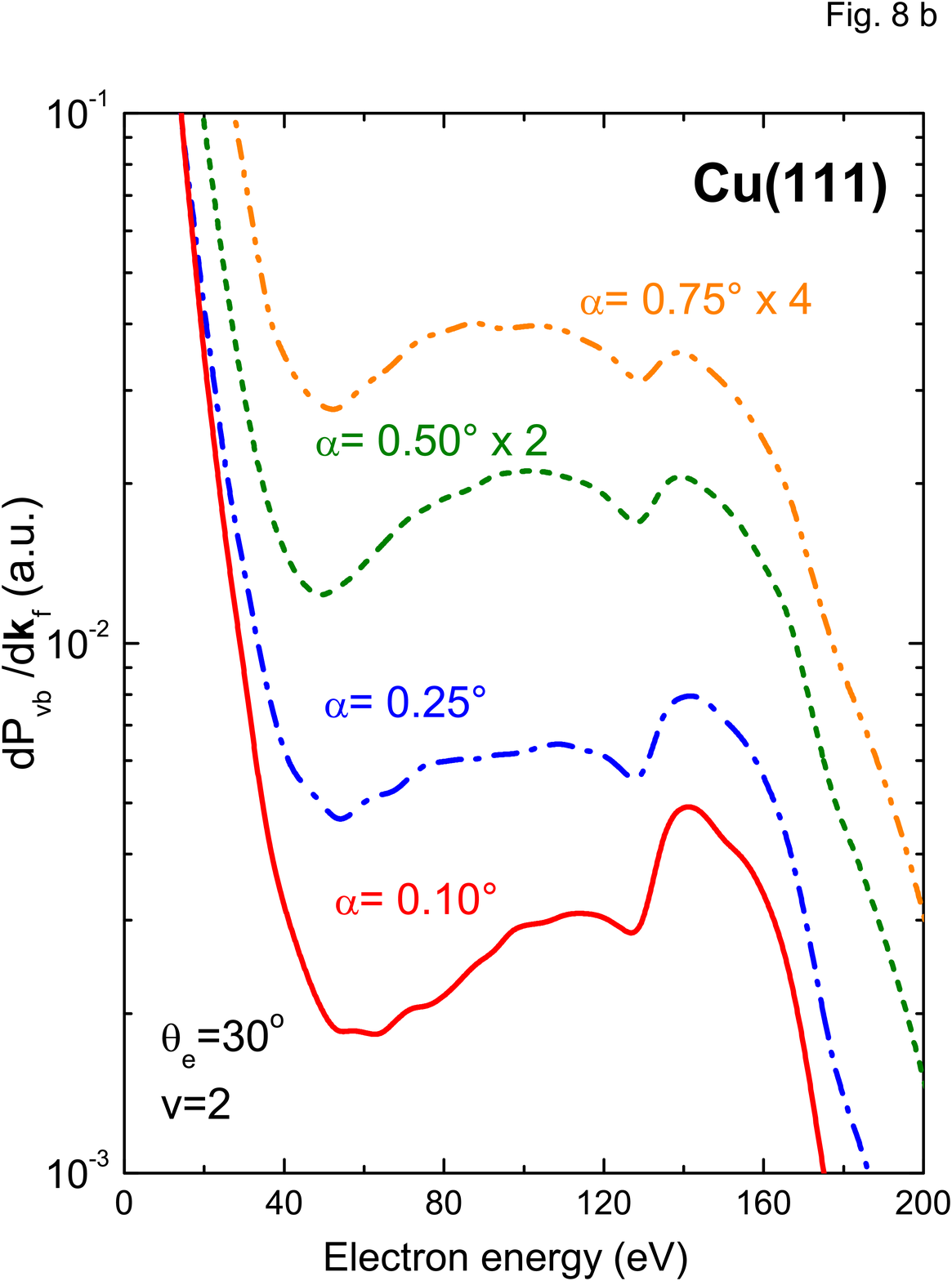}
\end{center}
\caption{(Color online) Influence of the incidence conditions on BSB
electron emission probabilities from the valence band of Cu(111).
The electron ejection angle is: $\protect\theta _{e}=30^{o}$. (a)
Different impact velocities, keeping the incidence angle,
$\protect\alpha =0.5^{o}$,
as a constant. (b) Different incidence angles, keeping the impact velocity, $%
v=2$ a.u., as a constant.}
\label{fig8}
\end{figure}


\begin{thebibliography}{99}
\bibitem{aclaracion} The corresponding initial states were described by
using Hartree-Fock wave functions for Mg$^{++}$, extracted from Ref. \cite%
{Clementi-Roetti}, pg. 332.
\end{thebibliography}
\end{document}